\documentclass[letterpaper,12pt]{article}
\usepackage[utf8]{inputenc}
\usepackage{graphicx}
\usepackage{subfigure}
\usepackage{bm}
\usepackage[margin=25mm]{geometry}
\usepackage{multirow}

\usepackage{amsmath}
\usepackage{amsfonts,amssymb}
\usepackage{mathrsfs}
\usepackage{theorem}
\usepackage{float}
\usepackage[round]{natbib}
\usepackage{epstopdf}

\def\realset{{\mathbb R}}
\def\Iset{{\mathbb I}}
\def\Pr{{\textnormal{Pr}}}

\def\flink{g}
\def\Enorm{\textnormal{E}}
\def\d{\textnormal{d}}

\newtheorem{Theorem}{Theorem}

\newtheorem{Proposition}{Proposition}
{\theorembodyfont{\rmfamily} }

\numberwithin{equation}{section}

\begin{document}

\begin{center}
{\Huge Categorical data analysis using a skewed Weibull regression model}

\vspace{0.3cm}
{\bf Renault Caron$^1$, Debajyoti Sinha$^2$, Dipak Dey$^3$, Adriano Polpo$^1$}\\
\vspace{-0.2cm}
$^1$ Department of Statistics, Federal University of S\~{a}o Carlos, Brazil\\
\vspace{-0.2cm}
$^2$ Department of Statistics, Florida State University, USA\\
\vspace{-0.2cm}
$^3$ College of Liberal Arts and Sciences, University of Connecticut, USA\\

\let\thefootnote\relax\footnote{Address for correspondence: Adriano Polpo, Universidade Federal de S\~{a}o Carlos, Departamento de Estat\'{i}stica, Rod. Washington Luiz, km 235, CEP: 13565-905, S\~{a}o Carlos/SP, Brazil. E-mail: polpo@ufscar.br}



\end{center}

\begin{abstract}
In this paper, we present a Weibull link (skewed) model for categorical response data arising from binomial as well as multinomial model. We show that, for such types of categorical data, the most commonly used models (logit, probit and complementary log-log) can be obtained as limiting cases. We further compare the proposed model with some other  asymmetrical models. The Bayesian as well as frequentist estimation procedures for binomial and multinomial data responses are presented in details. The analysis of two data sets to show the efficiency of the proposed model is performed.
\end{abstract}



\section{Introduction}
\label{sec_intro}

The statistical problem of estimating binary response variables is very important  in many areas including social science, biology and economics \citep{Agresti2009}. The vast bibliography of categorical data presents the big evolution of the methods that handle appropriately binary and polychotomous data. More details can be found in \citet{Agresti2002}. Generalized linear model (GLM) has a wide range of tools in regression for count data \citep{Nelder1972}. Two important and commonly used symmetric link functions in GLM are the logit and probit links \citep{McCullagh1989}. Many studies have investigated the limitations of these symmetric link functions. It is well accepted that when the probability of the binary response approaches 0 at a different rate from the rate (as a function of covariate) it approaches 1, symmetric link functions cannot be appropriate \citep{Chen1999c}. Many parametric classes of link functions are in the literature, including the power transform of logit link by \citet{Aranda-Ordaz1981} and the skew-normal link of \citet{Chen1999c}. Other works with one-parameter class include \citet{Guerrero1982,Morgan1983,Whittmore1983} and a host of others. Existing models for two-parameter families include \citet{Stukel1988,Prentice1976,Pregibon1980,Czado1992a} and \citet{Czado1994a}.

Stukel's model with transformation of both tails of logit link is very general and can approximate many important links including probit, logit and complementary log-log. However, the Bayesian analysis of Stukel's model is not straightforward to implement, particularly in presence of multiple covariates and noninformative improper priors. Skew-normal model via latent variable approach \citep{Albert1993} is convenient for sampling from the posterior distribution. However the frequentist analysis for the model proposed in \citet{Chen1999c} is not trivial. The existence of the maximum likelihood estimator (MLE) of the linear regression parameters ($\bm{\beta}$) can be proved only under the restrictive condition that the skewness parameter of the link function is known \citep{Bazan2010}.

The majority of the works in literature are devoted to the models for binary response data. For the case of multinomial data, the multinomial extension of the logit link \citep[chap. 8]{Hosmer2000} and associated inference tools are simple to perform, and the marginal distribution of each component preserves the logit link. As mentioned before, the symmetric link may not be always appropriate. We are also not aware of any model with asymmetric link function for multinomial data.

\citet{Caron2009} briefly suggested an asymmetrical link function, called Weibull link, exclusively for binary response data. In this paper, we take the Bayesian route and extend their work to multinomial data. Further we present for the first time the associated Bayesian inference tools and explore the properties of the proposed link function. We show that the benefits of this model are as follows: (1) flexibility of the Weibull distribution, (2) logit, probit and complementary log-log links as limiting cases, (3) case of implementation of both frequentist and Bayesian inferences, and (4) a general extension to handle multinomial response. The implementation of the associated Markov chain Monte Carlo (MCMC) algorithm to sample from posterior distribution is not complicated. Also, we develop an Empirical Bayes tool \citep{Robbins1956,Carlin2000} to obtain the prior when there is no relevant prior information available to the statistician. We illustrate the use of Weibull link via analysis of two following data examples. (1) For the experiment to study the potencies of three poisons \citep{Finney1947}, main binary response is whether the insect is alive after being treated with assigned dose level. For this example, we compare our Weibull link model with other asymmetric and symmetric link models. (2) Main response of  \citet{Grazeffe2008} study is the multiple levels of DNA damage in circulating hemocytes of each adult snail irradiated with an assigned dose. This study is used to illustrate the analysis of multinomial response data under Weibull link model, and comparing the results 
with those obtained by \citet{Grazeffe2008} using logistic regression.

The article is organized as follows. In Section \ref{sec_weib} we present the Weibull model, its novel properties and some approximations of the link function. In Section \ref{sec_est}, we present the estimation procedures using MLE as well as the Bayesian estimation. In Section \ref{sec_est}, we also present the estimation procedure for multinomial response. Section \ref{sec_ex} is devoted to illustrate the Weibull link for analyzing two real data sets, and comparison with other existing models. Finally, Section \ref{sec_final} presents some future considerations and final comments.

\section{Weibull regression model}
\label{sec_weib}

\subsection{Link function}
\label{sec_weib-link}

Let $\bm{X} = (\bm{1}, \bm{X_1}, \ldots, \bm{X_r})'$ be the design matrix, where $\bm{1}$ is a vector with all values equal to 1, $j=1, \ldots, r$. We denote the vector of binary response variable as $\bm{Y}$. Similar to GLM, our interest lies in modeling the probability $\Pr[Y_i = 1 \mid \eta_i] = \mu(\eta_i) = \Enorm(Y_i)$ as $\Pr[Y_i = 1 \mid \eta_i] = \flink^{-1}(\eta_i)$, $i=1,\ldots,n$, where $\bm{\eta} = \bm{\beta}\bm{X}$, $\bm{\beta} = (\beta_0, \beta_1, \ldots \beta_r)$ are the linear coefficients, and $\flink(\cdot)$ is the link function. The link function relates the covariates $\bm{X}$ with the mean response $\mu = \Enorm(Y \mid X)$. In this case, the $\flink^{-1}$ is a cumulative distribution function (cdf) on the real line. Our interest is a link function that can accommodate symmetric and asymmetric tails which has a simple parametric fuctional form, and can be easily tractable. To obtain these goals, we use the cdf of Weibull distribution 
\begin{equation}
\label{eq_weibull}
F(\eta) = 1-\exp\{-(\eta-\alpha)^\gamma\} \Iset_{(\eta > \alpha)},
\end{equation}
for $\flink^{-1}$, where $\alpha \in \realset$ is the location/threshold parameter, $\gamma > 0$ is the shape parameter, and $\Iset_{(\eta > \alpha)}$ is the indicator of $\eta > \alpha$. $\Iset_A$ is the indicator function of event $A$, that is $\Iset_{A} = 1$ and $\Iset_{A^c} = 0$.

Alternatively, the Weibull link function is defined as
\begin{equation}
\label{eq_weib-link}
\begin{array}{lll}
\eta = & g(\mu) & = \left[-\log\left(1-\mu \right) \right]^{\frac{1}{\gamma}}, \\
\mu = & g^{-1}(\eta) & = 1-\exp\{-\eta^\gamma \},
\end{array}
\end{equation}
where $\mu(\eta) = \Enorm(Y \mid \eta) \geq 0$, $\gamma > 0$, and $\eta > 0$.

\begin{figure}[!ht]
\centering
\includegraphics[width=0.5\textwidth]{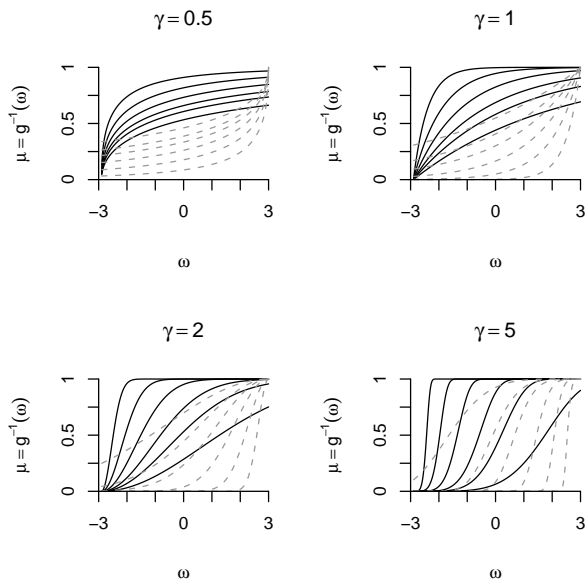}
\caption{Forms of Weibull link.}
\label{fig_form-weib}
\end{figure}

Note that in the above parametrization the restriction of $\eta > 0$ is not a problem because the parameter $\beta_0$ plays the role of both the location/threshold parameter $\alpha$ and the intercept of linear predictor $\bm{\eta} = \bm{\beta}\bm{X}$. By doing this, we avoid the identifiability problem in estimation of $\beta_0$, also we have a more parsimonious model. The skewness of the Weibull link depends only on the parameter $\gamma$, and can be evaluated by $(\Gamma_3 -  3\Gamma_2\Gamma_1 + 2\Gamma_1^3)/(\Gamma_2 - \Gamma_1^2)^{3/2}$, where $\Gamma_j = \Gamma(1+j/\gamma)$ and $\Gamma(\cdot)$ is the Gamma function. The skewness lies in the interval $(-1.1395,\infty)$. We also evaluated the Arnold-Groeneveld (AG) skewness measure \citep{ArnoldGroeneveld1995}, which is a skewness measure related to the mode of a distribution. Again, the AG skewness depends only on the parameter $\gamma$, and can be evaluated as $2\exp\{(1-\gamma)/\gamma\} -1$, and lies in the interval $(-0.26424, \infty)$. However, sometimes a model with skewness lower than $-1.1395$ is desired, in this case we can use the reflected Weibull distribution to define the link as $\mu = g^{-1}(\omega) = \exp\{-\eta^\gamma\}$, and the skewness lies in the interval $(-\infty,1.1395)$. The different forms of Weibull link are shown in Figure \ref{fig_form-weib} with solid line for the Weibull link and dashed line for the reflected Weibull link.

\subsection{Special cases}
\label{sec_weib-spec}

\begin{figure}[!ht]
\centering
\subfigure{\includegraphics[width=0.40\textwidth,keepaspectratio=true]{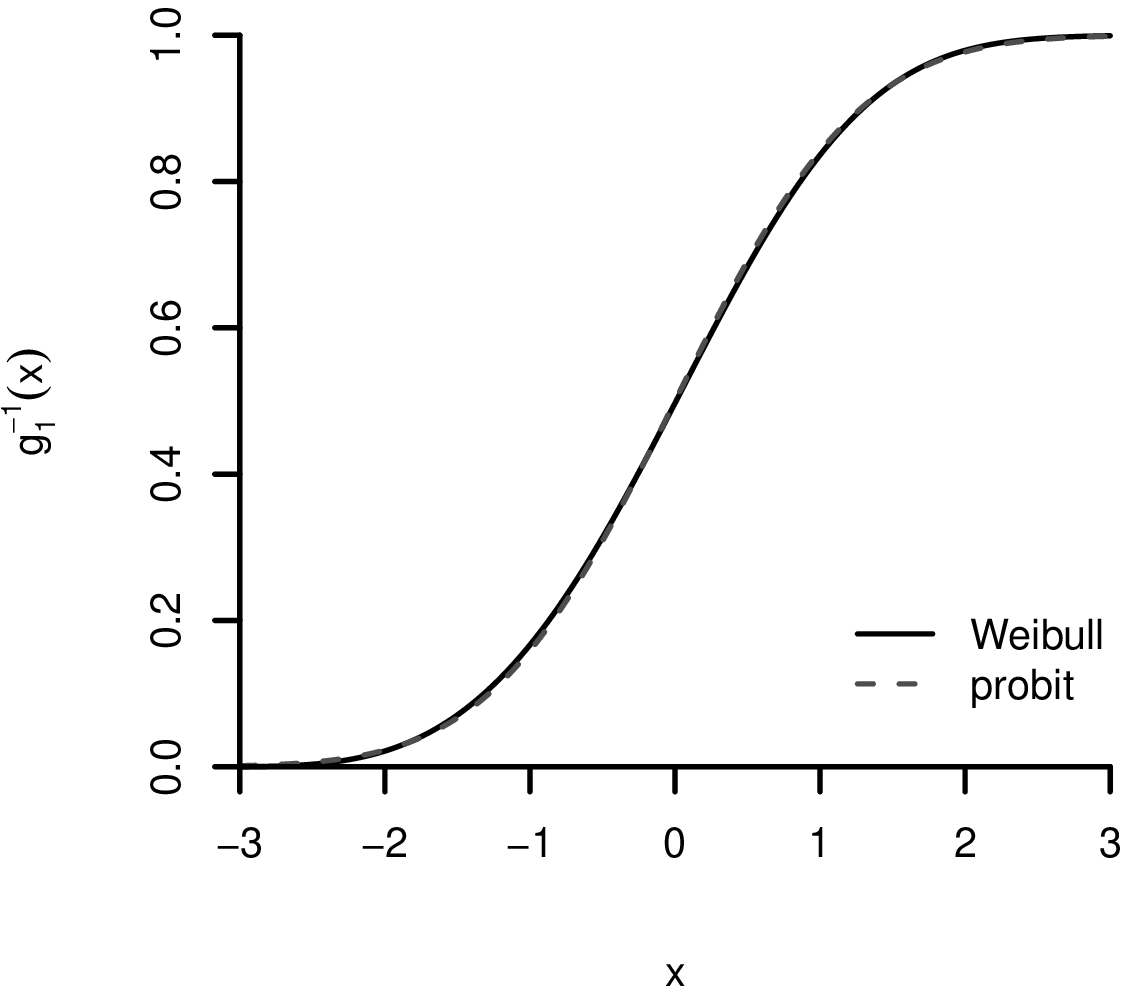}
\label{fig_approx-probit}}
\subfigure{\includegraphics[width=0.40\textwidth,keepaspectratio=true]{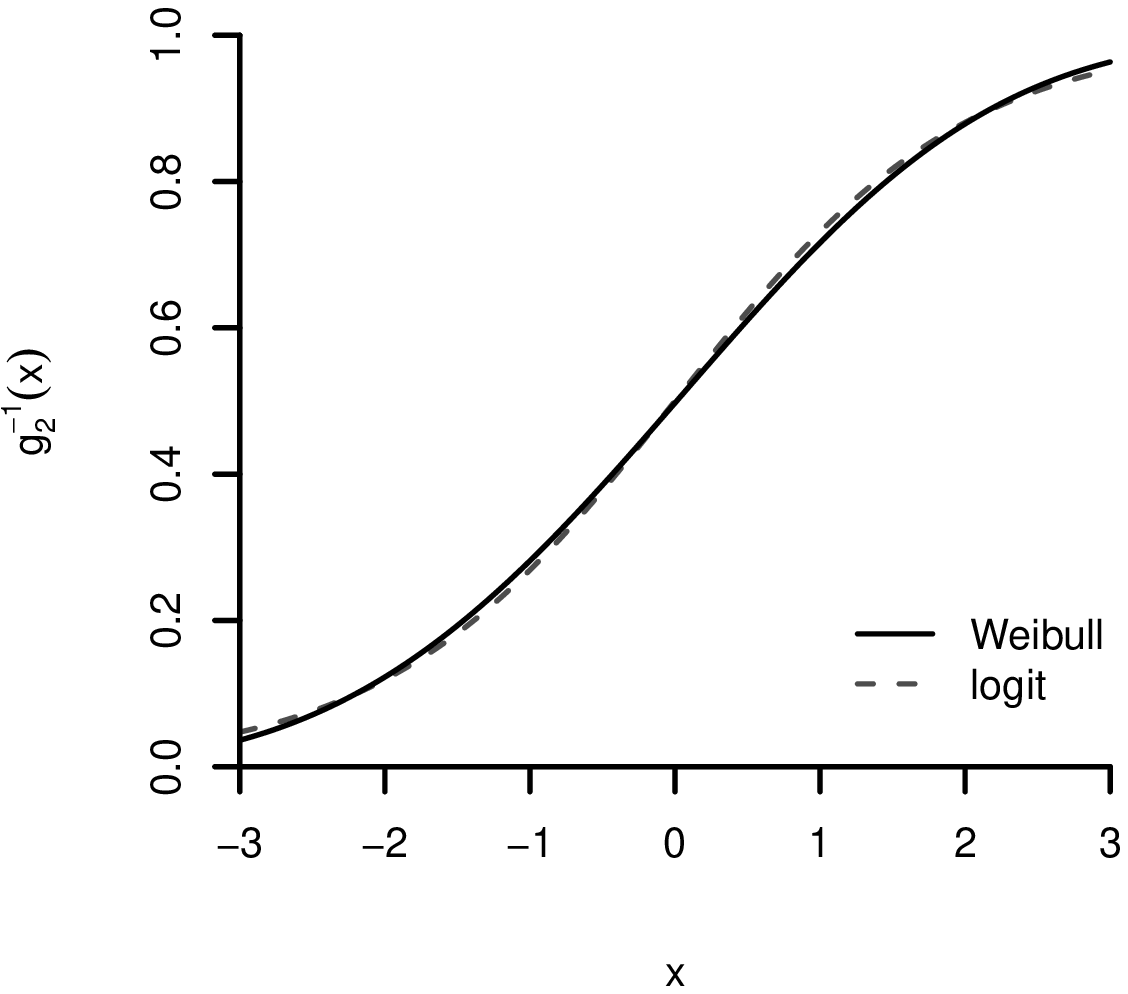}
\label{fig_approx-logit}}
\caption{Similarity of Weibull link with \subref{fig_approx-probit} probit link; and 
\subref{fig_approx-logit} logit link.}
\label{fig_apl}
\end{figure}

The choice of the Weibull distribution as link function is due to its flexible properties. \citet[Chapter 3]{Rinne2009} discusses the various properties of Weibull along with Weibull distribution as approximation to some symmetrical distributions. We highlight the relations of Weibull with the normal and logistic distributions, because they explain the relations of Weibull link with probit and logit link functions. Based on results of \citet{Rinne2009}, we have
\begin{eqnarray*}
g_1^{-1}(\eta) & = & 1-\exp\left\{-(0.90114+0.27787 \eta)^{3.60235} \right\} \approx \Phi(\eta), \\
g_2^{-1}(\eta) & = & 1-\exp\left\{-(0.89864+0.16957 \eta)^{3.50215} \right\} \approx \frac{\exp(\eta)}{1+\exp(\eta)},
\end{eqnarray*}
where $\Phi$ is the distribution function of the standard normal distribution. These results show that Weibull link can approximate the probit link and the logit link. The degrees of these approximations are illustrated in Figure \ref{fig_apl}.

We have the following proposition for another important case of link, the complementary log-log link \citep{McCullagh1989}.

\begin{Proposition}
\label{prop_cll}
The complementary log-log link defined by $g^{-1}(\eta) = 1-\exp\{-\exp(\eta)\}$ is a limiting case of the Weibull link because
\begin{equation}
\label{eq_limit-cll}
\lim\limits_{\gamma \rightarrow \infty} 
   \left\{ 1-\exp\left[-\left(1+ 
       \frac{\eta}{\gamma}\right)^\gamma \right] \right\} 
    = 1-\exp\{-\exp(\eta)\}.
\end{equation}
\end{Proposition}

Taking $\alpha = -1$ in (\ref{eq_weibull}) and dividing $\eta$ by $\gamma$, without loss of generality, we can rewrite the Weibull link given in (\ref{eq_weib-link}) as 
\[ g^{-1}(\eta) = 1-\exp\left\{-\left(1+ \frac{\eta}{\gamma}\right)^\gamma \right\}. \]
Now, taking the limit $\gamma \rightarrow \infty$ of $g^{-1}(\eta)$ completes the proof.

\begin{figure}[!ht]
\centering
\subfigure{\includegraphics[width=0.40\textwidth,keepaspectratio=true]{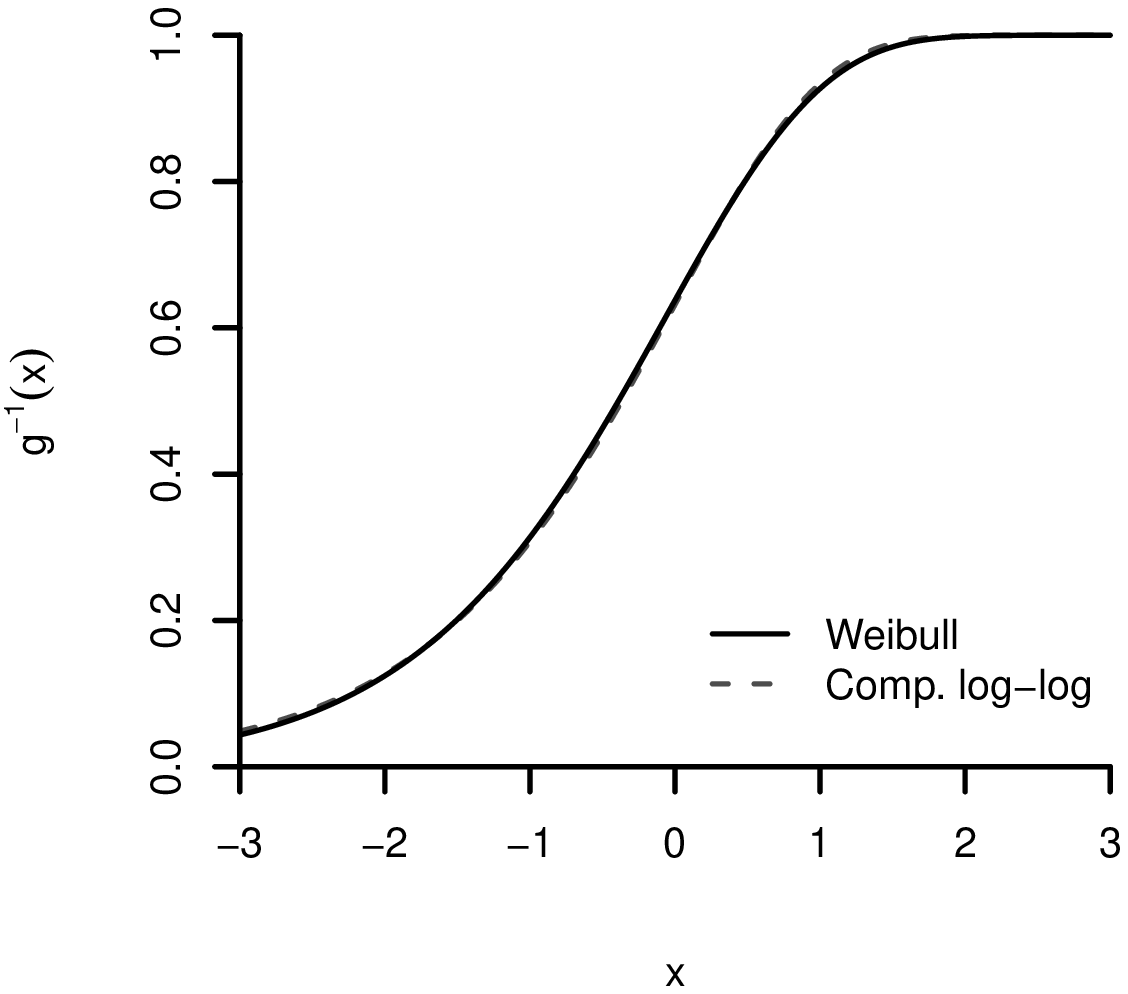}
\label{fig_approx-cll}}
\subfigure{\includegraphics[width=0.40\textwidth,keepaspectratio=true]{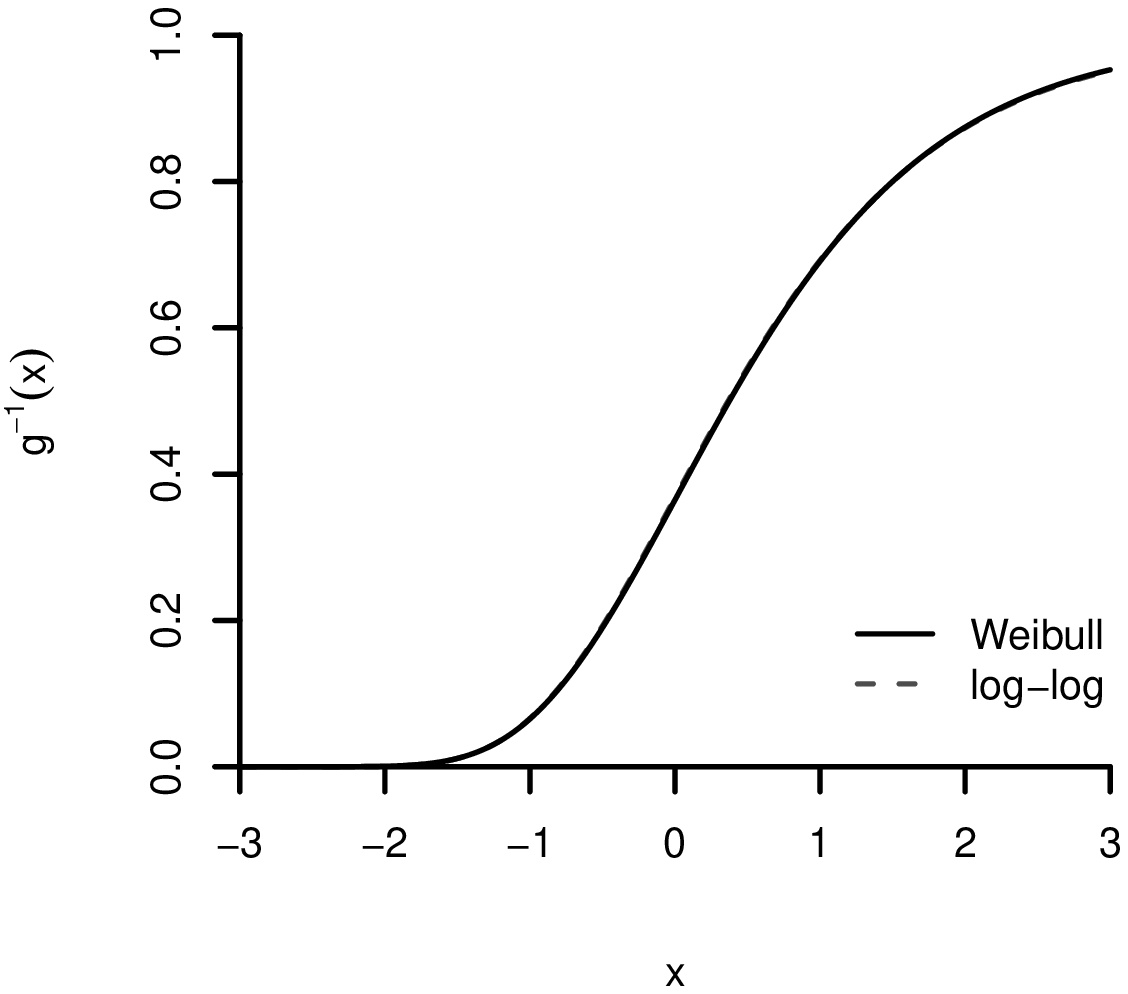}
\label{fig_approx-ll}}
\caption{Similarity of Weibull link with \subref{fig_approx-cll} complementary log-log link; and 
\subref{fig_approx-ll} log-log link.}
\label{fig_ap-cll-ll}
\end{figure}

Given this result, we can say that for a data set when the estimated value of $\gamma$ is large then the complementary log-log link should be appropriate. Using the reflected Weibull link, we have a similar result with the log-log link, defined as $g^{-1}(\eta) = \exp\{-\exp(-\eta)\}$ \citep{McCullagh1989}. The complementary log-log and log-log link as limiting cases are illustrated in Figure \ref{fig_ap-cll-ll}.

\section{Estimation}
\label{sec_est}

\subsection{Binomial data}
\label{sec_est-bin}

For now, consider a sample of size $n$ from the binary variable/response $Y$, with  \mbox{$\Pr[Y_i =1] =$} $p_i$ for $i=1, \ldots, n$. We denote the observed data as ${\cal D} = \{n,\bm{Y} = \bm{y},\bm{X} = \bm{x}\}$, where $\bm{y} = (y_1, \ldots, y_n)$ is the observed vector of $\bm{Y} = (Y_1, \ldots, Y_n)$, and $\bm{x} = (\bm{1}, \bm{x_1}, \ldots, \bm{x_r})'$, is the observed covariate matrix of $\bm{X} = (\bm{1}, \bm{X_1}, \ldots, \bm{X_r})'$. The likelihood function for the Weibull link can be written as
\begin{eqnarray}
\label{eq_lik}
\hspace*{-0.4cm} L\left(\bm{\beta},\gamma \mid {\cal D} \right) 
  &\hspace*{-0.25cm} \propto & \hspace*{-0.25cm}
    \prod\limits_{i=1}^{n} {p_i}^{y_i} (1-p_i)^{1-y_i} \nonumber \\
  &\hspace*{-0.25cm} \propto & \hspace*{-0.25cm} \prod\limits_{i=1}^{n} 
    \left[1 - \exp\left\{-\eta_i^\gamma \right\} \right]^{y_i}
    \hspace*{-0.1cm} \left[ \exp\left\{ -\eta_i^{\gamma}
                             \right\} \right]^{1-y_i}\hspace*{-0.1cm},
\end{eqnarray}
\noindent
and the log-likelihood as
\begin{equation}
\label{eq_log-lik}
l\left(\bm{\beta},\gamma \mid {\cal D} \right) \propto \sum\limits_{i=1}^{n}
     \left[ y_i \log\left\{1 - \exp\left(-\eta_i^\gamma \right) \right\}
                        - (1-y_i) \eta_i^{\gamma} \right],
\end{equation}
where $\eta_i$ is the $i$-th element of the vector $\bm{\eta} = \bm{\beta}\bm{X}$, and $\bm{\beta}$, $\gamma$ are the parameters to be estimated.

A numerical method such as \citet{Nelder1965} can be used to obtain the MLE for $(\bm{\beta}, \gamma)$. The expression of the gradient vector and Hessian matrix are given in Appendix \ref{sec_ap-grad-hess}. As initial guesses for the numerical algorithm, we suggest using the estimator $\widetilde{\bm{\beta}}_{i,probit}$ under probit model for $\beta_{i}$ ($i \neq 0$), $\widetilde{\beta}_{0,guess} = -\min(\widetilde{\bm{\beta}}_{probit} \bm{x})+0.001$ for $\beta_0$, and $3.60235$ for $\gamma$. The initial guesses $(\bm{\beta}, \gamma)$ can be interpreted as the Weibull link being an approximate probit link.

For the Bayesian analysis, the posterior density is
\begin{equation}
\label{eq_post}
p(\bm{\beta}, \gamma \mid {\cal D}) \propto L(\bm{\beta}, \gamma \mid {\cal D}) p(\bm{\beta}, \gamma),
\end{equation}
where $p(\bm{\beta}, \gamma)$ is the joint prior. We suggest using the hierarchical Bayes model. Assuming the parameters to a priori independent, the first level of hierarchy has $\gamma$ following a gamma distribution with mean $m_\gamma$ and variance $v_\gamma$, and $\bm{\beta}$ with multivariate normal distribution with mean vector $m_{\bm{\beta}}$ and covariance matrix $v_{\bm{\beta}} \bm{I}$, where $\bm{I}$ is the identity matrix. The values of $v_\gamma$ and $v_{\bm{\beta}}$ are fixed, and for $m_\gamma$ and $m_{\bm{\beta}}$ we consider a uniform (improper) prior, that is $p(m_{\gamma}) = p(m_{\bm{\beta}}) \propto 1$. Arguably, we can use the mode of the integrated likelihood of ($m_\gamma$, $m_{\bm{\beta}}$) to determine a prior distribution \citep{Carlin2000}. The hyper-parameters $v_\gamma$ and $v_{\bm{\beta}}$ are viewed as prior precision parameters. The EM algorithm \citep{McLachlan2008} can be used to obtain the estimates of $m_\gamma$ and $m_{\bm{\beta}}$. The MCMC procedure is used to generate a sample from the posterior distribution. We are omitting the details about these computational tools because they are already well known tools and are not the main subject of the present paper.

Another advantage of the Weibull link is that the posterior distributions are proper even when we use a wide range of non-informative priors. The Jeffreys' prior for the parameter $\bm{\beta}$ has the form $p(\bm{\beta} \mid \gamma) \propto |\bm{I}(\bm{\beta} \mid \gamma)|^{1/2}$, where the Fisher information matrix $\bm{I}(\bm{\beta} \mid \gamma)$ can be obtained by taking the expectation of the Hessian matrix given in Appendix \ref{sec_ap-grad-hess}.

Considering the uniform prior $p(\bm{\beta}) \propto 1$, and the non-informative prior $p(\gamma) \propto 1/\gamma^c$, for $\gamma > 1$ and $c > 1$ a known constant \citep{Sun1997}, we have the non-informative prior distribution 
\begin{equation}
\label{eq_prior}
p(\bm{\beta}, \gamma) \propto p(\bm{\beta})p(\gamma) \propto \frac{1}{\gamma^c}.
\end{equation}
With this constraint (in the parameter $\gamma$ of the Weibull link) the skewness lies in the interval $(-1.1395,2]$, which still are a flexible link. For the improper prior of (\ref{eq_prior}), the propriety of the resulting posterior distribution in (\ref{eq_post}) is stated in Theorem \ref{teo_post}.

\begin{Theorem}
\label{teo_post}
Let $z_i = -1$ when $y_i =0$ and $z_i = 1$ when $y_i = 1$, and $\bm{X}^*$ be the matrix with rows $z_i \bm{x}_i'$. Suppose that the design matrix $\bm{X}$ is of full rank, and there exists a positive vector $\bm{a} = (a_1, \ldots, a_n)' \in \realset^n$, with $a_i > 0$, for $i=1, \ldots, n$, such that ${\bm{X}^*}' \bm{a} = 0$, under the non-informative prior of (\ref{eq_prior}), then the posterior density (\ref{eq_post}) is proper.
\end{Theorem}

{\it Proof.} Let $u, u_1, \ldots, u_n$ be independent random variables with common Weibull distribution with shape parameter $\gamma$. For $0 < k < \infty$ we have that $\Enorm(|u|^k) = \Gamma(1 + k/\gamma) < \infty$. Observing that $1-F(x) = \Enorm[\Iset(u > x)]$ and $F(x) = \Enorm[\Iset(u \leq x)]$, where $\Iset$ is an indicator function. Then, we have $[F(\bm{x}_i' \bm{\beta})]^{y_i} [1-F(\bm{x}_i'\bm{\beta})]^{1-y_i} \leq \Enorm(z_i u_i \geq z_i\bm{x}_i' \bm{\beta})$ and $[F(\bm{x}_i' \bm{\beta})]^{y_i} [1-F(\bm{x}_i'\bm{\beta})]^{1-y_i} \geq \Enorm(z_iu_i > z_i\bm{x}_i' \bm{\beta})$. Let $\bm{u}^* = (z_1 u_1, \ldots, z_n u_n)$. By the Fubini's theorem, we get
\begin{eqnarray}
&&\hspace{-1.5cm} \int_1^\infty \int_{\realset^k} L(\bm{\beta},\gamma \mid \bm{y} \bm{X}) \frac{1}{\gamma^c} \d \bm{\beta} \d \gamma \nonumber \\
& = & \int_1^\infty \frac{1}{\gamma^c} \int_{\realset^n} \int_{\realset^k} \Iset(z_i u_i > z_i \bm{x}_i' \bm{\beta}, 1 \leq i \leq n) \d \bm{\beta} \d \bm{F}(\bm{u}) \d \gamma \nonumber \\
& = & \int_1^\infty \frac{1}{\gamma^c} \int_{\realset^n} \int_{\realset^k} \Iset(\bm{X}^* \bm{\beta} \leq \bm{u}^*) \d \bm{\beta} \d \bm{F}(\bm{u}) \d \gamma. \nonumber
\end{eqnarray}
From Lemma 4.1 of \citet{Chen2000} there exists a constant $K$ depending only on $\bm{X}^*$ such that 
\[ \int_{\realset^k} \Iset(\bm{X}^* \bm{\beta} \leq \bm{u}^*) \d \bm{\beta} \leq K ||\bm{u}^*||^k, \]
which yields
\[ \int_1^\infty \int_{\realset^k} L(\bm{\beta},\gamma \mid \bm{y} \bm{X}) \frac{1}{\gamma^c} \d \bm{\beta} \d \gamma < \infty, \]
by $\Enorm(|u|^k) < \infty$, and $\int_1^\infty 1/\gamma^c \d \gamma < \infty$ for $c > 1$.
\begin{flushright}
$\blacksquare$
\end{flushright}

\subsection{Multinomial data}
\label{sec_est-mult}

For multinomial responses, we have that $Y_i \in \{1, \ldots, {\cal K}\}$, and \mbox{$p_k = \Pr(Y_i = k)$}, for $k=1,\ldots,{\cal K}$ and $\sum_{j=1}^{\cal K} p_j = 1$. The likelihood function for multinomial response data $\mathcal{D}$ is
\begin{equation}
L\left(\bm{p} | {\cal D} \right) \propto \prod\limits_{i = 1}^n \prod\limits_{k = 1}^{\cal K} p_k^{\Iset(y_i = k)} = \prod\limits_{k = 1}^{\cal K} p_k^{s_k},
\end{equation}
where $s_k = \sum\limits_{i=1}^n \Iset(y_i = k)$, and $\Iset_A$ is the indicator function of event $A$, that is $\Iset_{A} = 1$ and $\Iset_{A^c} = 0$. Note that $\sum\limits_{k=1}^{\cal K} s_k = n$.

Using a reparameterization \citep{Pereira2008b} of $\bm{p}$ as $p_1 = \theta_1$, $p_k = \theta_k \prod\limits_{\ell=1}^{k-1} (1-\theta_\ell)$, for $k = 1, \ldots, \mathcal{K}-1$, and $p_{\mathcal{K}} = \prod\limits_{\ell=1}^{\mathcal{K}-1} (1-\theta_\ell)$ the likelihood function can be rewritten as
\begin{equation}
L\left(\bm{p} | {\cal D} \right) \propto \prod\limits_{k=1}^{{\cal K}-1}
\theta_k^{s_k} (1-\theta_k)^{n-\sum\limits_{\ell = 1}^{k} s_\ell} = 
\prod\limits_{k=1}^{{\cal K}-1} L\left(\theta_k | {\cal D} \right).
\end{equation}
This shows that the estimation for multinomial data is equivalent to estimating binomial response models. For MLE, we have $\widehat{p}_1 = \widehat{\theta}_1$, $\widehat{p}_k = \widehat{\theta}_k \prod\limits_{\ell=1}^{k-1} (1-\widehat{\theta}_\ell)$, and $\widehat{p}_{\cal K} = \prod\limits_{\ell=1}^{{\cal K}-1} (1-\widehat{\theta}_\ell)$. For Bayesian estimation, we generate a sample from the posterior distribution of each $\theta_k$, then we can do the transformation to obtain the estimators of $\bm{p}$. For this, we need to generate a sample from the posterior of $\gamma_k$ and $\bm{\beta}_k$, for each $k = 1, \ldots, \mathcal{K}-1$, and then perform the proper transformation to obtain the sample from the posterior of $\theta_k$. In this case, the prior of $\theta_k$ can be viewed as a transformation of the priors of $\gamma_k$ and $\bm{\beta}_k$. And so, for both MLE and Bayesian estimator, we can use the procedures described in Section \ref{sec_est-bin}. The partition scheme presented to solve the multinomial model estimation is intuitive. For more details about the reparameterization used here see \citep{Pereira2008b}.

\subsection{Model selection and diagnostics}
\label{sec_est-}

In the case of binomial data, to compare models within frequentist set up, we use the Akaike Information Criterion -- AIC \citep{Akaike1974} and the Bayesian Information Criterion -- BIC \citep{Schwarz1978}. For Bayesian analysis, we use long established  tool of Deviance Information Criterion -- DIC \citep{Spiegelhalter2002}. We omit the details 
of these popular tools for the sake of brevity. Also, we use a version of Kolmogorov-Smirnov statistics (KS) as measure of goodness of fit. KS is defined as $\textnormal{KS} = \max_i |y_i - \widehat{y_i}|$, the maximum absolute error of the predicted and the observed frequencies, where $\widehat{y_i}$ is the predicted value of $y_i$. We also use Mean Absolute Error (MAE) defined as $\textnormal{MAE} = \frac{1}{n} \sum_{i=1}^n |y_i - \widehat{y_i}|$.

\section{Data Example}
\label{sec_ex}

\subsection{Binary data example}
\label{sec_ex-bin}

We analyze the study of relative potency of three different poisons: Rotenone, Deguelin and Mixture \citep[][p. 69]{Finney1947}. The experiment was to test the different poisons with different doses, with objective to understand the potency of the poisons. It was considered 5 doses for rotenone, 6 doses for deguelin and 6 doses for the mixture. For each dose and poison it was considered around 50 insects, and observed how many insects was killed. The data are presented in Table \ref{tab_ex-bin_data}. We consider that the response variable is binary with $Y=1$ representing the insect killed, and as covariates: $X_1$ as the log(Dose), $X_2$ as an indicator of Rotonone, and $X_3$ as an indicator of Deguelin. The mixture of poisons is considered as the reference poison (that is, $X_2 = 0$ and $X_3 = 0$). Our main objective is to find the model that better represent (fit) these Data. We are not looking for the ``best'' poison or dose.

\begin{table}[!ht]
\caption{\label{tab_ex-bin_data} Relative potency of Rotenone, a Deguelin concentrate, and a Mixture of two.}
\centering
\begin{tabular}{ccc|ccc|ccc}
\hline
\multicolumn{3}{c}{Rotenone} & 
\multicolumn{3}{|c|}{Deguelin} & 
\multicolumn{3}{c}{Mixture} \\
\hline
log(Dose) & dead & $n$ & log(Dose) & dead & $n$ & log(Dose) & dead & $n$\\
\hline
1.01 & 44 & 50 & 1.70 & 48 & 48 & 1.40 & 48 & 50 \\
0.89 & 42 & 49 & 1.61 & 47 & 50 & 1.31 & 43 & 46 \\
0.71 & 24 & 46 & 1.48 & 47 & 49 & 1.18 & 38 & 48 \\
0.58 & 16 & 48 & 1.31 & 34 & 48 & 1.00 & 27 & 46 \\
0.41 & ~6 & 50 & 1.00 & 18 & 48 & 0.71 & 22 & 46 \\
     &    &    & 0.71 & 16 & 49 & 0.40 & ~7 & 47 \\
\hline
\multicolumn{9}{l}{{\footnotesize Table from \citet[][p. 69]{Finney1947}.}}
\end{tabular}
\end{table}

\begin{figure}[!ht]
\centering
\subfigure{\includegraphics[width=0.32\textwidth,keepaspectratio=true]{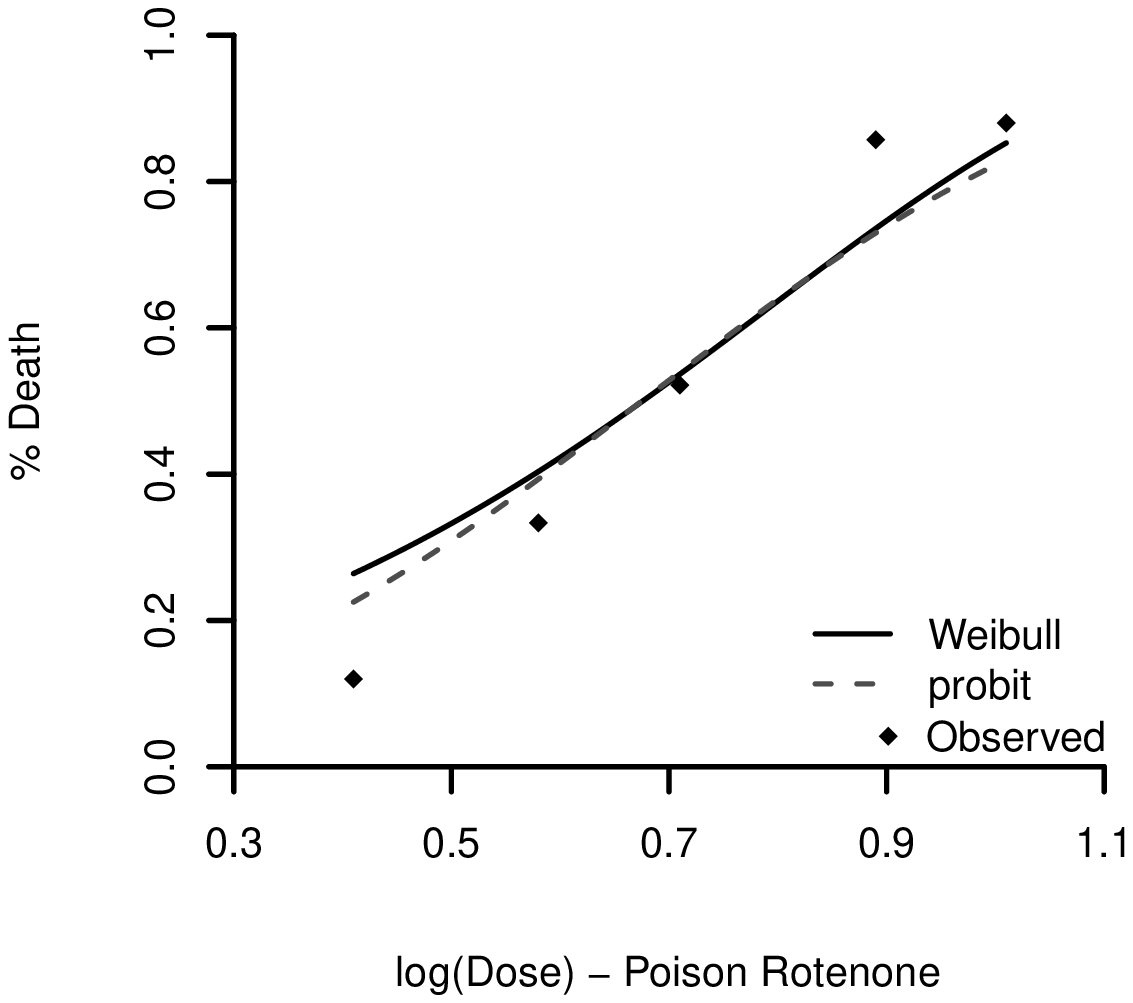}
\label{fig_mle-bin_rotonone}}
\subfigure{\includegraphics[width=0.32\textwidth,keepaspectratio=true]{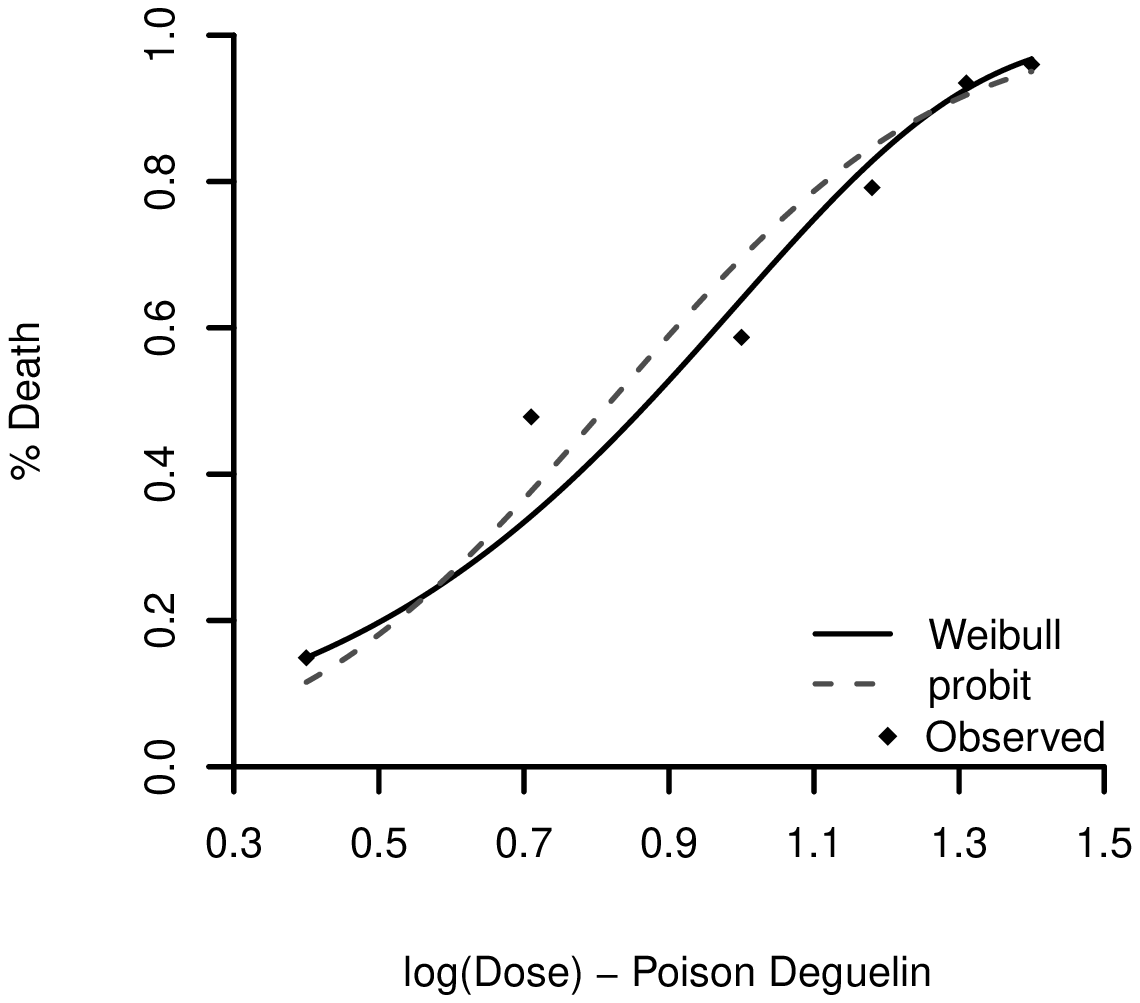}
\label{fig_mle-bin_deguelin}}
\subfigure{\includegraphics[width=0.32\textwidth,keepaspectratio=true]{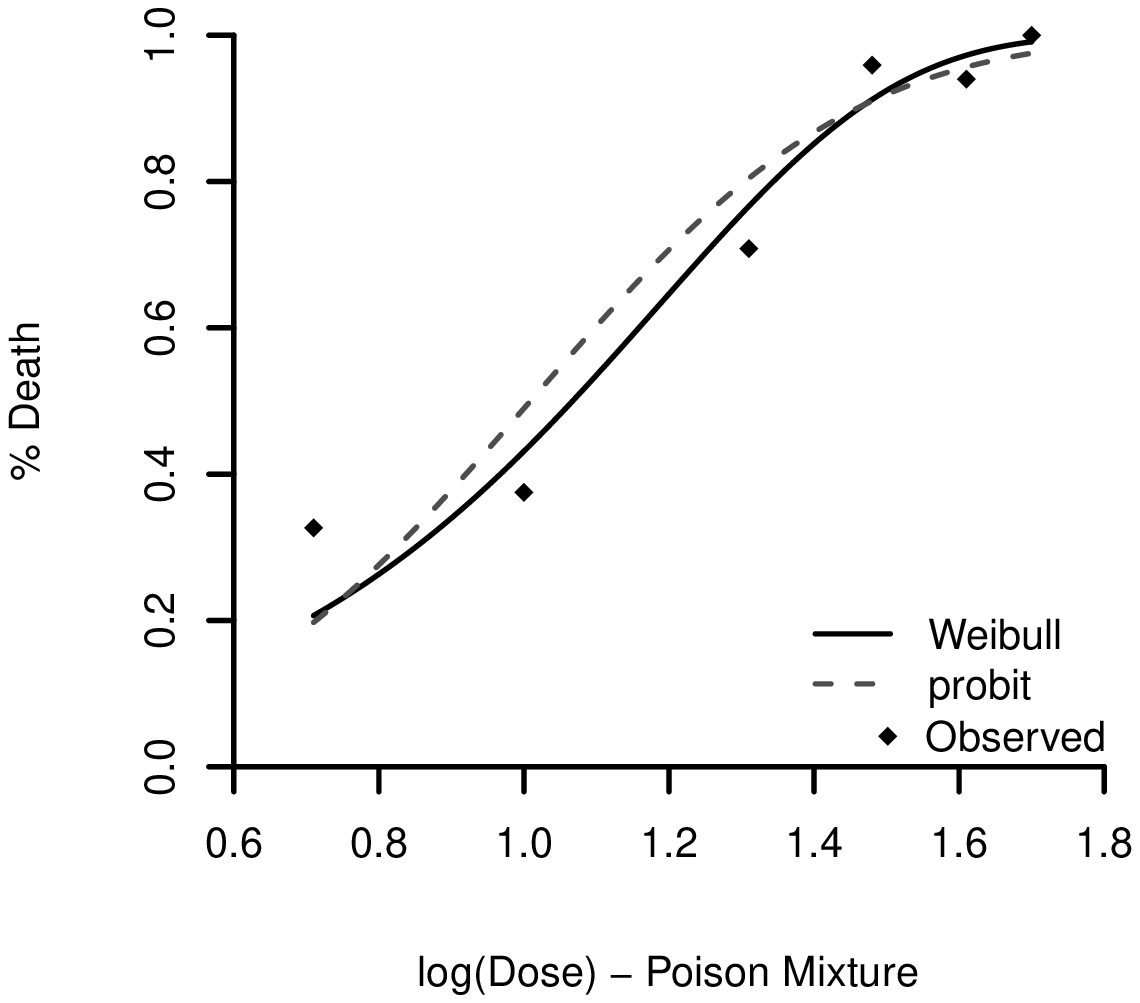}
\label{fig_mle-bin_mixture}}
\caption{Comparison of MLE for Weibull and probit link for three types of poisons: 
\subref{fig_mle-bin_rotonone} Rotonone; \subref{fig_mle-bin_deguelin} Deguelin; and 
\subref{fig_mle-bin_mixture} Mixture.}
\label{fig_mle-bin}
\end{figure}

We obtain the MLE for Weibull parameters and for comparison we also estimated the parameters of complementary log-log, Stukel, probit, logit, Aranda-Ordaz, and Prentice models. Table \ref{tab_ex-bin_mle-comp}, presents some statistics of each model to compare them. The best models are complementary log-log and Weibull. The advantage of the complementary log-log is that this model has one fewer parameter. However, as expected the Weibull model performs similar to the complementary log-log (see Proposition \ref{prop_cll}). The model with lower KS is the probit model. As we observe in Figure \ref{fig_mle-bin}, the Weibull model has a better fit than the probit model. The estimated parameter values of Weibull and probit model are given in Table \ref{tab_ex-bin-mle_est}.

\begin{table}[!ht]
\caption{\label{tab_ex-bin_mle-comp} Comparison of the link functions under MLE.}
\centering
\begin{tabular}{cccccc}
\hline
Model       & $\log(L)$ &    AIC &    BIC &     KS &    MAE \\
\hline
comp. log-log & -370.33 & 748.66 & 767.48 & 0.1451 & 0.0551 \\
Weibull       & -370.34 & 750.69 & 774.22 & 0.1440 & 0.0553 \\
Stukel        & -369.51 & 751.01 & 779.26 & 0.1704 & 0.0481 \\
Prentice      & -369.66 & 751.32 & 779.56 & 0.1583 & 0.0492 \\
probit        & -372.57 & 753.14 & 771.97 & 0.1292 & 0.0656 \\
logit         & -373.41 & 754.82 & 773.65 & 0.1351 & 0.0668 \\
Aranda-Ordaz  & -374.90 & 759.80 & 783.33 & 0.3674 & 0.1396 \\
\hline
\end{tabular}
\end{table}

\begin{table}[!ht]
\caption{\label{tab_ex-bin-mle_est} MLE for the binomial example.}
\centering
\begin{tabular}{ccc}
\hline
& \multicolumn{2}{c}{Model [estimate (SE)]} \\
Parameter  & Weibull & probit \\
\hline
$\beta_0$ & ~~0.9735 (~0.0110) & -2.3364 (0.1953) \\
$\beta_1$ & ~~0.0266 (~0.0111) & ~2.8478 (0.1832) \\
$\beta_2$ & ~~0.0053 (~0.0024) & ~0.4138 (0.1333) \\
$\beta_3$ & ~-0.0051 (~0.0024) & -0.5369 (0.1367) \\
$\gamma$  & 114.5084 (47.9818) & --- \\
\hline
\end{tabular}
\end{table}

Another important model is the skew-probit proposed by \citet{Chen1999c}, and to compare with the Weibull model we perform a Bayesian analysis for both models. The priors for the parameters of the Weibull model are the same as that described in Section \ref{sec_est-bin}. We use the values $v_\gamma = 100$ and $v_{\bm{\beta}} = 25$, which is equivalent to set a large variance in the priors and can be viewed as a prior with less information. The estimated values for the hyper-parameters of first hierarchical level are $\widehat{m_\gamma} = 9.1125$ and $\widehat{\bm{m_\beta}} = (0.1548, 0.9029, 0.1273, -0.1619)$. For the priors of skew-probit model, we used a uniform distribution over the interval $(-1,1)$ for the asymmetry parameter and independent normal distribution with mean $0$ and variance $25$ for each $\beta_j$, $j = 0, \ldots, 3$, which also is a less informative prior. Table \ref{tab_ex-bin_be-comp} presents the model selection criteria (DIC, KS and MAE) for two competing models. For everyone of three criteria, a smaller value indicates a better agreement between the Weibull model and observed data. We note that the Weibull model has better fit than the skew-probit model. For both Weibull and skew-probit models, the posterior means of relevant parameters are given in Table \ref{tab_ex-bin-be_est}.

\begin{table}[!ht]
\caption{\label{tab_ex-bin_be-comp} Comparison of the link functions under Bayesian estimation.}
\centering
\begin{tabular}{cccc}
\hline
            &    DIC &     KS &    MAE \\
\hline
Weibull     & 753.43 & 0.1305 & 0.0660 \\
skew-probit & 755.72 & 0.1462 & 0.0667 \\
\hline
\end{tabular}
\end{table}

\begin{table}[!ht]
\caption{\label{tab_ex-bin-be_est} Bayesian estimates for binomial example.}
\centering
\begin{tabular}{cccc}
\hline
          & \multicolumn{2}{c}{Model [posterior mean (standard deviation)]} \\
Parameter & Weibull & skew-probit \\
\hline
$\beta_0$ & ~0.1342 (0.1692) & -2.1617 (0.4322)\\
$\beta_1$ & ~0.9276 (0.1983) & ~2.6283 (0.2666)\\
$\beta_2$ & ~0.1284 (0.0436) & ~0.3842 (0.1238)\\
$\beta_3$ & -0.1727 (0.0593) & -0.4942 (0.1321)\\
$\gamma$ (or $\delta$)  & ~3.2420 (0.7326) & -0.0058 (0.4795)\\
\hline
\end{tabular}
\end{table}

\subsection{Multinomial data example}
\label{sec_ex-mult}

\citet{Grazeffe2008} reported a study of DNA mutation of the cells of adult snails, each irradiated with a single dose of gamma radiation. They recorded four categories of DNA mutation with $Y = 1, 2, 3$ and $4$ representing no mutation, low, intermediate, and  high DNA mutation respectively. The snails are randomized into 5 different dose levels with $X \in \{0, 2.5, 5, 10, 20\}$. The data are presented in Table \ref{tab_ex-mult_data}. The objective is to compare effects of different dose levels on DNA mutation $Y$. We illustrate the use of Weibull link model for analysis of this study with multinomial responses. Further, in Table \ref{tab_ex-mult}, we compare our estimates of $\Pr[Y = k \mid x]$ with those obtained by \citet{Grazeffe2008} based on the logit link model.

\begin{table}[!ht]
\caption{\label{tab_ex-mult_data} Comet assay: DNA damage in hemocytes of B. glabrata exposed to 60 Co gamma radiation.}
\centering
\begin{tabular}{c|cccc|c}
\hline
& \multicolumn{4}{|c|}{DNA damage classes} & \\
Dose(Gy) & $Y = 1$ & $Y = 2$ & $Y = 3$ & $Y = 4$ & Number of cells\\
\hline
$0$   & 654 & 125 & ~72 & 249 & 1100\\
$2.5$ & 442 & 178 & 105 & 175 & ~900\\
$5$   & 197 & 253 & 173 & 277 & ~900\\
$10$  & 159 & 296 & 264 & 281 & 1000\\
$20$  & ~58 & ~49 & 133 & 660 & ~900\\
\hline
\multicolumn{6}{l}{{\footnotesize Table from \citet{Grazeffe2008}.}}
\end{tabular}
\end{table}

For a proper comparison with previous method of \citet{Grazeffe2008}, we obtain the MLE with only $X$ and $X^2$ as covariates. We use the reflected Weibull link, because this model has lower values of KS and MAE than those for Weibull link. To obtain the estimation of the reflected Weibull model we first estimate the values of $\theta_1$, $\theta_2$, $\theta_3$. To simplify, consider the three binary variables $Z_1$, $Z_2$ and $Z_3$, where $\theta_k = \Pr(Z_k = k)$, $k = 1, 2, 3$. Then, using the results is Section \ref{sec_est-mult} and the data in Table \ref{tab_ex-mult_data} we constructed the Table \ref{tab_ex-mult_Z} with the observed values of $Z$'s, and estimate the models for $Z$'s.

\begin{table}[!ht]
\caption{\label{tab_ex-mult_Z} Observed values of constructed variables $Z_1$, $Z_2$ and $Z_3$.}
\centering
\begin{tabular}{c|cc|cc|cc}
\hline
         & \multicolumn{2}{c|}{$Z_1$} & \multicolumn{2}{c|}{$Z_2$} &
                                        \multicolumn{2}{c}{$Z_3$} \\
Dose ($X$) & $0$ & $1$ & $0$ & $1$ & $0$ & $1$ \\
\hline
     0   & 446 & 654 & 321 & 125 & 249 &  72 \\
     2.5 & 458 & 442 & 280 & 178 & 175 & 105 \\
     5   & 703 & 197 & 450 & 253 & 277 & 173 \\
    10   & 841 & 159 & 545 & 296 & 281 & 264 \\
    20   & 842 &  58 & 793 &  49 & 660 & 133 \\
\hline
\end{tabular}
\end{table}

The parameter estimates for the three binary models are presented in Table \ref{tab_ex-mult_est}, and we have $\widehat{\theta_1}(\bm{x}) = e^{-(0.0234 +1.6395 x -0.6748 x^2)^{0.1742}}$, $\widehat{\theta_2}(\bm{x}) = e^{-(1.0930 +0.0368 x -0.0030 x^2)^{2.3604}}$ and $\widehat{\theta_3}(\bm{x}) = e^{-(1.2429 +0.0866 x -0.0047 x^2)^{1.7562}}$.

\begin{table}[!ht]
\caption{\label{tab_ex-mult_est} MLE of Weibull model for the multinomial example.}
\centering
\begin{tabular}{c|ccc}
\hline
& \multicolumn{3}{c}{Model [estimate (SE)]}\\
parameter & $Y = 1$ & $Y = 2$ & $Y = 3$ \\
\hline
$\gamma$  & ~0.1742 (0.0209) & ~2.3604 (0.1905) & ~1.7562 (0.1415)\\
$\beta_0$ & ~0.0234 (0.0123) & ~1.0930 (0.0258) & ~1.2429 (0.0369)\\
$\beta_1$ & ~1.6395 (0.8355) & ~0.0368 (0.0067) & ~0.0866 (0.0078)\\
$\beta_2$ & -0.6748 (0.3295) & -0.0030 (0.0002) & -0.0047 (0.0003)\\
\hline
\end{tabular}
\end{table}

\begin{figure}[!ht]
\centering
\includegraphics[width=0.7\textwidth,keepaspectratio=true]{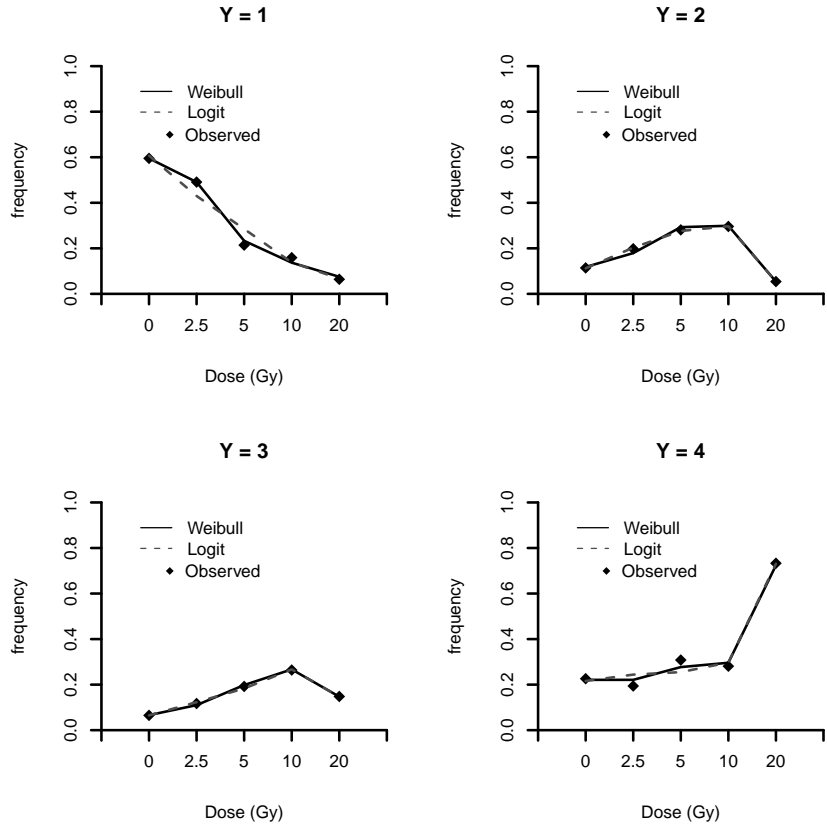}
\caption{Estimated population frequencies.}
\label{fig_ex-mult_est}
\end{figure}

As described in Section \ref{sec_est-mult}, we have that $\widehat{p_1}(\bm{x}) = \widehat{\theta_1}(\bm{x})$, $\widehat{p_2}(\bm{x}) =$ \mbox{$[1-\widehat{\theta_1}(\bm{x})] \widehat{\theta_2}(\bm{x})$}, $\widehat{p_3}(\bm{x}) = [1-\widehat{\theta_1}(\bm{x})] [1-\widehat{\theta_2}(\bm{x})] \widehat{\theta_3}(\bm{x})$, and $\widehat{p_4}(\bm{x}) =$ \mbox{$[1-\widehat{\theta_1}(\bm{x})] [1-\widehat{\theta_2}(\bm{x})] [1-\widehat{\theta_3}(\bm{x})]$}, where $\widehat{p_k}(\bm{x})$ is the estimated value of $\Pr[Y = k | x]$, $k = 1,\ldots, 4$. The estimated frequencies of DNA damage for each class is presented in Table \ref{tab_ex-mult}, and illustrated in Figure \ref{fig_ex-mult_est}. This figure shows that the Weibull link model has a better fit for categories $Y = 1$ and $4$. For categories $Y = 2$ and $ 3$, both models have comparable performances. Table \ref{tab_ex-mult_comp} presents the inferential statistics for model comparisons. Every one of these indicates a substantial preference for Weibull link model.

\begin{table}[!ht]
\caption{\label{tab_ex-mult_comp} Comparison of the link functions for multinomial example.}
\centering
\begin{tabular}{ccccc}
\hline
        &      AIC &    KS &    MAE \\
\hline
Weibull & 11332.83 & 0.031 & 0.0097 \\
logit   & 11362.39 & 0.071 & 0.0165 \\
\hline
\end{tabular}
\end{table}

\begin{table}[!ht]
\caption{\label{tab_ex-mult} Relative frequencies of damage (observed and model's estimates).}
\centering
\begin{tabular}{cc|cccc}
\hline
                     &          & \multicolumn{4}{c}{Damage classes ($Y$)} \\
Dose(Gy)             &  Model   &     1 &     2 &     3 &     4 \\
\hline
\multirow{3}{*}{0}   & Observed & 0.595 & 0.114 & 0.065 & 0.226 \\
                     & Weibull  & 0.595 & 0.120 & 0.065 & 0.220 \\
                     & Logit    & 0.607 & 0.112 & 0.065 & 0.216 \\
\hline
\multirow{3}{*}{2.5} & Observed & 0.491 & 0.198 & 0.117 & 0.194 \\
                     & Weibull  & 0.491 & 0.178 & 0.112 & 0.219 \\
                     & Logit    & 0.430 & 0.203 & 0.123 & 0.244 \\
\hline
\multirow{3}{*}{5}   & Observed & 0.214 & 0.281 & 0.192 & 0.308 \\
                     & Weibull  & 0.233 & 0.289 & 0.200 & 0.278 \\
                     & Logit    & 0.285 & 0.276 & 0.183 & 0.255 \\
\hline
\multirow{3}{*}{10}  & Observed & 0.159 & 0.296 & 0.264 & 0.281 \\
                     & Weibull  & 0.137 & 0.302 & 0.264 & 0.296 \\
                     & Logit    & 0.138 & 0.297 & 0.267 & 0.297 \\
\hline
\multirow{3}{*}{20}  & Observed & 0.064 & 0.054 & 0.148 & 0.733 \\
                     & Weibull  & 0.075 & 0.054 & 0.147 & 0.725 \\
                     & Logit    & 0.067 & 0.054 & 0.147 & 0.731 \\
\hline
\end{tabular}
\end{table}

\section{Final Comments}
\label{sec_final}

In this paper we have presented a Weibull model to estimate the problem of binary and multi\-no\-mial regression analysis. The model is very flexible and capable to handle with many different types of data. The comparison with other skew-link model, in binomial data example (Section \ref{sec_ex-bin}), shows that the performance of the Weibull link was better than all (equal to complementary log-log). We are convinced that our proposed model is a very good option, at least for our motivating study. A good feature of the model is that the logit, probit, complementary log-log, and log-log link functions are approximations of Weibull link. Then the proposed model can accommodate even symmetric link function. For this flexibility of the Weibull link model, we are comfortable to suggest its use in practice.

Other aspect of the proposed Weibull model is that the associated numerical procedure of MLE is very simple to implement, particularly in comparison to other competing. For Bayesian estimates we also suggested an Empirical Bayes approach to determine the prior. Under full Bayesian estimation, we compare the model with the skew-probit model proposed by \citet{Chen1999c}, in Section \ref{sec_ex-bin}. The performance of our model is better even under full Bayesian framework.

We also develop a partition scheme for the multinomial regression model simplifying the problem to ${\cal K}-1$ binomial regression analysis. This is a general scheme that can be used for other link functions, which opens a vast options to estimate multinomial data. Future works can include the comparison between different models, the goodness of fit statistics and other different models for multinomial data analysis.



\appendix

\section{Gradient and Hessian for log-likelihood of Weibull model}
\label{sec_ap-grad-hess}

Let $\xi_{i1} = \exp\{-\eta_i^\gamma\}$, $\xi_{i2} = \exp\{-2 \eta_i^\gamma\}$, $i=1,\ldots,n$.

\subsection{Gradient}

The gradient vector is $\bm{g} = (g_1, g_2, \ldots, g_r)$, where
\begin{eqnarray*}
g_1 = \frac{\partial l(\gamma,\bm{\beta} | {\cal D})}{\partial \gamma} & = &
 \sum\limits_{i=1}^n
 -(1-y_i) \eta_i^{\gamma} \log(\eta_i) 
 +\frac{y_i \xi_{i1} \eta_i^{\gamma} \log(\eta_i)}
       {1-\xi_{i1}};\\
\end{eqnarray*}
\begin{eqnarray*}
g_2 = \frac{\partial l(\gamma,\bm{\beta} | {\cal D})}{\partial \beta_0} & = &
 \sum\limits_{i=1}^n
 -\gamma (1-y_i) \eta_i^{\gamma-1}
 +\frac{\gamma y_i \xi_{i1} \eta_i^{\gamma-1}}
       {1-\xi_{i1}};
\end{eqnarray*}
for $j = 1, \ldots, r-2$,
\begin{eqnarray*}
g_{(j+2)} = \frac{\partial l(\gamma,\bm{\beta} | {\cal D})}{\partial \beta_j} & = &
 \sum\limits_{i=1}^n
 -\gamma x_{ij} (1-y_i) \eta_i^{\gamma-1}
 + \frac{\gamma x_{ij} y_i \xi_{i1} \eta_i^{\gamma-1}}{1-\xi_{i1}}.
\end{eqnarray*}

\subsection{Hessian matrix}

The Hessian matrix $\bm{H}$ is
\begin{equation*}
\bm{H} = \left[ \begin{array}{cccc}
h_{11} & h_{12} & \cdots & h_{1r} \\
h_{21} & h_{22} & \cdots & h_{2r} \\
\vdots & \vdots & \ddots & \vdots \\
h_{r1} & h_{r2} & \cdots & h_{rr}
\end{array} \right],
\end{equation*}
where, for $j = 1, \ldots, r-2$,
\begin{eqnarray*}
h_{11} = \frac{\partial^2 l(\gamma,\bm{\beta} | {\cal D})}{\partial \gamma \partial \gamma} & = & 
 \sum\limits_{i=1}^n
 -(1-y_i) [\log(\eta_i)]^2 \eta_i^{\gamma} \\
 &~& ~~~~~ +\frac{ \{\xi_{i1} y_i [\log(\eta_i)]^2\} [\eta_i^{\gamma} - \eta_i^{2 \gamma}]}
       {1-\xi_{i1}} \\
 &~& ~~~~~ -\frac{\xi_{i2} y_i [\log(\eta_i)]^2 \eta_i^{2 \gamma}}
       {\left(1-\xi_{i1}\right)^2};
\end{eqnarray*}

\begin{eqnarray*}
h_{21} = \frac{\partial^2 l(\gamma,\bm{\beta} | {\cal D})}{\partial \gamma \partial \beta_0} & = & 
 \frac{\partial^2 l(\gamma,\bm{\beta} | {\cal D})}{\partial \beta_0 \partial \gamma} = h_{12};
\end{eqnarray*}

\begin{eqnarray*}
h_{(j+2)1} = \frac{\partial^2 l(\gamma,\bm{\beta} | {\cal D})}{\partial \gamma \partial \beta_j} 
 & = & \frac{\partial^2 l(\gamma,\bm{\beta} | {\cal D})}
            {\partial \beta_j \partial \gamma} = h_{1(j+2)};
\end{eqnarray*}

\begin{eqnarray*}
h_{12} = \frac{\partial^2 l(\gamma,\bm{\beta} | {\cal D})}{\partial \beta_0 \partial \gamma} & = & 
 \sum\limits_{i=1}^n
 -[1 +\gamma \log(\eta_i)] (1-y_i) \eta_i^{\gamma-1} \\
 &~& ~~~~~ +\frac{[1+(1- \eta_i^{\gamma})\gamma \log(\eta_i)] \xi_{i1} y_i \eta_i^{\gamma-1}}
       {1-\xi_{i1}} \\
 &~& ~~~~~ -\frac{\xi_{i2} \gamma y_i \log(\eta_i) \eta_i^{2 \gamma-1}}
       {\left(1-\xi_{i1}\right)^2};
\end{eqnarray*}

\begin{eqnarray*}
h_{22} = \frac{\partial^2 l(\gamma,\bm{\beta} | {\cal D})}{\partial \beta_0 \partial \beta_0} & = & 
 \sum\limits_{i=1}^n
 -(\gamma-1) \gamma (1-y_i) \eta_i^{\gamma-2} \\
 &~& ~~~~~ +\frac{[(\gamma-1) -\gamma \eta_i^\gamma] \xi_{i1} \gamma y_i \eta_i^{\gamma-2}}
       {1-\xi_{i1}} \\
 &~& ~~~~~ -\frac{\xi_{i2} \gamma^2 y_i \eta_i^{2 \gamma-2}}
       {\left(1-\xi_{i1}\right)^2};
\end{eqnarray*}

\begin{eqnarray*}
h_{(j+2)2} = \frac{\partial^2 l(\gamma,\bm{\beta} | {\cal D})}{\partial \beta_0 \partial \beta_j} 
 & = & \sum\limits_{i=1}^n
 -(\gamma-1) \gamma x_{ij} (1-y_i) \eta_i^{\gamma-2} \\
 &~& ~~~~~ +\frac{[(\gamma-1) -\gamma \eta_i^\gamma] \xi_{i1} \gamma x_{ij} y_i \eta_i^{\gamma-2}}
       {1-\xi_{i1}} \\
 &~& ~~~~~ -\frac{\xi_{i2} \gamma^2 x_{ij} y_i \eta_i^{2 \gamma-2}}
       {\left(1-\xi_{i1}\right)^2};
\end{eqnarray*}

\begin{eqnarray*}
h_{1(j+2)} = \frac{\partial^2 l(\gamma,\bm{\beta} | {\cal D})}{\partial \beta_j \partial \gamma} 
 & = & \sum\limits_{i=1}^n
 -[x_{ij} (1-y_i) \eta_i^{\gamma-1}] [1+\gamma \log(\eta_i)] \\
 &~& ~~~~~ +\frac{\xi_{i1} x_{ij} y_i \eta_i^{\gamma-1} [1 + \gamma \log(\eta_i) (1 - \eta_i^{\gamma})]}
       {1-\xi_{i1}} \\
 &~& ~~~~~ -\frac{\xi_{i2} \gamma x_{ij} y_i \log(\eta_i) \eta_i^{2 \gamma-1}}
       {\left(1-\xi_{i1}\right)^2};
\end{eqnarray*}

\begin{eqnarray*}
h_{2(j+2)} = \frac{\partial^2 l(\gamma,\bm{\beta} | {\cal D})}{\partial \beta_j \partial \beta_0} 
 & = & \frac{\partial^2 l(\gamma,\bm{\beta} | {\cal D})}
            {\partial \beta_0 \partial \beta_j} = h_{(j+2)2};
\end{eqnarray*}
for $k = 1, \ldots, r$ and $k \neq j$,
\begin{eqnarray*}
h_{(k+2)(j+2)} = \frac{\partial^2 l(\gamma,\bm{\beta} | {\cal D})}
                       {\partial \beta_j \partial \beta_k} & = &
 \sum\limits_{i=1}^n
 -(\gamma-1) \gamma x_{ij}x_{ik} (1-y_i) \eta_i^{\gamma-2} \\
 &~& ~~~~~ +\frac{[(\gamma-1) - \gamma \eta_i^{\gamma}] \xi_{i1} \gamma x_{ij}x_{ik} y_i \eta_i^{\gamma-2}}
       {1-\xi_{i1}} \\
 &~& ~~~~~ -\frac{\xi_{i2} \gamma^2 x_{ij}x_{ik} y_i \eta_i^{2 \gamma-2}}
       {\left(1-\xi_{i1}\right)^2} = h_{(j+2)(k+2)}.
\end{eqnarray*}


\begin{thebibliography}{}

\bibitem[Agresti, 2002]{Agresti2002}
Agresti, A. (2002).
\newblock {\em Categorical Data Analysis}.
\newblock John Wiley \& Sons, 2nd edition.

\bibitem[Agresti and Finlay, 2009]{Agresti2009}
Agresti, A. and Finlay, B. (2009).
\newblock {\em Statistical Methods for the Social Sciences}.
\newblock Pearson Prentice Hall, 4th edition.

\bibitem[Akaike, 1974]{Akaike1974}
Akaike, H. (1974).
\newblock A new look at the statistical model identification.
\newblock {\em IEEE Transactions on Automatic Control}, 19(6):716--723.

\bibitem[Albert and Chib, 1993]{Albert1993}
Albert, J.~H. and Chib, S. (1993).
\newblock {B}ayesian analysis of binary and polychotomous response data.
\newblock {\em Journal of the American Statistical Association}, 88:669--679.

\bibitem[Aranda-Ordaz, 1981]{Aranda-Ordaz1981}
Aranda-Ordaz, F.~J. (1981).
\newblock On two families of transformations to additivity for binary response
  data.
\newblock {\em Biometrika}, 68:357--363.

\bibitem[Arnold and Groeneveld, 1995]{ArnoldGroeneveld1995}
Arnold, B.~C. and Groeneveld, R.~A. (1995).
\newblock Measuring skewness with respect to the mode.
\newblock {\em The American Statistician}, 49:34--38.

\bibitem[Baz\'{a}n et~al., 2010]{Bazan2010}
Baz\'{a}n, J.~L., Bolfarine, H., and Branco, M.~D. (2010).
\newblock A framework for skew-probit links in binary regression.
\newblock {\em Communications in Statistics - Theory and Methods}, 39:678--697.

\bibitem[Carlin and Louis, 2000]{Carlin2000}
Carlin, B.~P. and Louis, T.~A. (2000).
\newblock {\em {B}ayes and Empirical {B}ayes Methods for Data Analysis}.
\newblock Chapman \& Hall/CRC, 2nd edition.

\bibitem[Caron, 2010]{Caron2010}
Caron, R. (2010).
\newblock Binary data regression: {W}eibull distribution (in portuguese).
\newblock Master's thesis, Department of Statistics -- Federal University of
  S\~{a}o Carlos, S\~{a}o Carlos, Brazil.

\bibitem[Caron and Polpo, 2009]{Caron2009}
Caron, R. and Polpo, A. (2009).
\newblock Binary data regression: {W}eibull distribution.
\newblock {\em AIP Conference Proceedings}, 1193(1):187--193.

\bibitem[Chen et~al., 1999]{Chen1999c}
Chen, M.-H., Dey, D.~K., and Shao, Q.-M. (1999).
\newblock A new skewed link model for dichotomous quantal response data.
\newblock {\em Journal of the American Statistical Association},
  94(448):1172--1186.

\bibitem[Chen and Shao, 2000]{Chen2000}
Chen, M.-H. and Shao, Q.-M. (2000).
\newblock Property of posterior distribution for dichotomous quantal response
  models with general link functions.
\newblock In {\em Proceedings of the American Mathematical Society}, volume
  129, pages 293--302.

\bibitem[Czado, 1992]{Czado1992a}
Czado, C. (1992).
\newblock {\em Advances in GLIM and Statistical Modelling}, volume~78 of {\em
  Lecture Notes in Statistics}, chapter On Link Selection in Generalized Linear
  Models.
\newblock Springer-Verlag.

\bibitem[Czado, 1994]{Czado1994a}
Czado, C. (1994).
\newblock Parametric link modification of both tails in binary regression.
\newblock {\em Statistical Papers}, 35:189--201.

\bibitem[Finney, 1947]{Finney1947}
Finney, D.~J. (1947).
\newblock {\em Probit Analysis}.
\newblock University Press, Cambridge.

\bibitem[Grazeffe et~al., 2008]{Grazeffe2008}
Grazeffe, V.~S., Tallarico, L.~F., Pinheiro, A.~S., Kawano, T., Suzuki, M.~F.,
  Okazaki, K., Pereira, C. A.~B., and Nakano, E. (2008).
\newblock Establishment of the comet assay in the freshwater snail biomphalaria
  glabrata (say, 1818).
\newblock {\em Mutation Research/Genetic Toxicology and Environmental
  Mutagenesis}, 654:58--63.

\bibitem[Guerrero and Johnson, 1982]{Guerrero1982}
Guerrero, V.~M. and Johnson, R.~A. (1982).
\newblock Use of the {B}ox-{C}ox transformation with binary response models.
\newblock {\em Biometrika}, 69:309--314.

\bibitem[Hosmer and Lemeshow, 2000]{Hosmer2000}
Hosmer, D.~W. and Lemeshow, S. (2000).
\newblock {\em Applied Logistic Regression}.
\newblock John Wiley \& Sons, New York, 2nd edition.

\bibitem[McCullagh and Nelder, 1989]{McCullagh1989}
McCullagh, P. and Nelder, J.~A. (1989).
\newblock {\em Generalized Linear Models}.
\newblock Chapman and Hall, London, 2nd edition.

\bibitem[McLachlan and Krishnan, 2008]{McLachlan2008}
McLachlan, G.~J. and Krishnan, T. (2008).
\newblock {\em The {EM} Algorithm and Extensions}.
\newblock John Wiley \& Sons.

\bibitem[Morgan, 1983]{Morgan1983}
Morgan, B. J.~T. (1983).
\newblock Observations on quantitative analysis.
\newblock {\em Biometrics}, 39:879--886.

\bibitem[Nelder and Mead, 1965]{Nelder1965}
Nelder, J.~A. and Mead, R. (1965).
\newblock A simplex algorithm for function minimization.
\newblock {\em Computer Journal}, 7:308--313.

\bibitem[Nelder and Wedderburn, 1972]{Nelder1972}
Nelder, J.~A. and Wedderburn, R. (1972).
\newblock Generalized linear models.
\newblock {\em Journal of the Royal Statistical Society, Ser. A}, 135:370--384.

\bibitem[Pereira and Stern, 2008]{Pereira2008b}
Pereira, C. and Stern, J. (2008).
\newblock Special characterizations of standard discrete models.
\newblock {\em {REVSTAT} - Statistical Journal}, 6(3):199--230.

\bibitem[Pregibon, 1980]{Pregibon1980}
Pregibon, D. (1980).
\newblock Goodness of link test for generalized linear models.
\newblock {\em Applied Statistics}, 29:338--345.

\bibitem[Prentice, 1976]{Prentice1976}
Prentice, R. (1976).
\newblock Generalization of the probit and logit models.
\newblock {\em Biometrics}, 32:761--768.

\bibitem[Rinne, 2009]{Rinne2009}
Rinne, H. (2009).
\newblock {\em The {W}eibull Distribution: A Handbook}.
\newblock CRC Press.

\bibitem[Robbins, 1956]{Robbins1956}
Robbins, H. (1956).
\newblock An empirical {B}ayes approach to statistics.
\newblock In {\em Proccedings of Third Berkeley Symposium on Mathematical
  Statistics and Probability}, volume~1, pages 157--163.

\bibitem[Schwarz, 1978]{Schwarz1978}
Schwarz, G.~E. (1978).
\newblock Estimating the dimension of a model.
\newblock {\em Annals of Statistics}, 6:461--464.

\bibitem[Spiegelhalter et~al., 2002]{Spiegelhalter2002}
Spiegelhalter, D.~J., Best, N.~G., Carlin, B.~P., and van~der Linde, A. (2002).
\newblock {B}ayesian measures of model complexity and fit.
\newblock {\em Journal of the Royal Statistical Society, Series B},
  64:583--639.

\bibitem[Stukel, 1988]{Stukel1988}
Stukel, T. (1988).
\newblock Generalized logistic models.
\newblock {\em Journal of the American Statistical Association},
  83(402):426--431.

\bibitem[Sun, 1997]{Sun1997}
Sun, D. (1997).
\newblock A note on noninformative priors for {W}eibull distributions.
\newblock {\em Journal of Statistical Planning and Inference}, 61:319--338.

\bibitem[Whittmore, 1983]{Whittmore1983}
Whittmore, A.~S. (1983).
\newblock Transformation to linearity in binary regression.
\newblock {\em SIAM Journal on Applied Mathematics}, 43:703--710.

\end{thebibliography}

\end{document}